\documentclass[12pt]{article}
\usepackage{amssymb,amsmath,epsfig}

\begin{document}

\title{\bf Phantom energy accretion onto a black hole in Ho$\check{r}$ava Lifshitz gravity }

\author{G.
Abbas \thanks{Email: ghulamabbas@ciitsahiwal.edu.pk}
\\
Department of Computer Science, COMSATS Institute of \\Information
Technology Sahiwal, Pakistan}
\date{}
\maketitle
\begin{abstract}
In this Letter, we examine the phantom energy accretion onto a
Kehagias-Sfetsos black hole in Ho$\check{r}$ava Lifshitz gravity. To
discuss the accretion process onto the black hole, the equations of
phantom flow near the black hole have been derived. It is found that
mass of the black hole decreases because of phantom accretion. We
discuss the conditions for critical accretion. Graphically, it has
been found that the critical accretion phenomena is possible for
different values of parameters. The results for the Schwarzschild
black hole can be recovered in the limiting case.
\end{abstract}

\
\

\section{Introduction}

Currently, observational cosmology has revealed that our universe is
in an accelerating phase. It has been verified by the data of
type-Ia Supernova and a large-scale structure [1-4]. Further, the
anisotropic behavior of radiations in cosmic microwave background
(CMB) as predicted by WMAP [5-7] favor the accelerating expansion of
universe. The exotic energy with negative pressure, known as
\textit{dark energy} (DE), is responsible for accelerating behavior
of the universe. Despite several observational facts, the nature of
DE is still an open issue in physics.

It is well-known that when phantom energy from an external source
accretes onto BH, then mass of BH decreases such that it eventually
attains extremal state and finally converts to NS. During accretion
process, charge and angular momentum remain unchanged. In Newtonian
theory, the problem of accretion of matter onto the compact object
was formulated by Bondi [8]. Michel [9] derived the relativistic
formula for the accretion of perfect fluid onto the Schwarzschild
BH. Babichev et al. [10] investigated that phantom accretion onto a
BH that can decrease its the mass if the back reaction effects of
accreting phantom fluid on geometry of BH are neglected. Jamil et
al. [11] discussed the critical accretion on the RN BH. They
determined the mass to charge ratio beyond which a BH can be
converted into a NS. The same conclusion was drawn by Babichev et
al. [12], by using the linear EoS and Chaplygin gas EoS for RN BH.
Madrid and Gonzalez [13] showed that accreting phantom energy onto
Kerr BH can convert it into a NS. Sharif and Abbas [14-16] discussed
the phantom energy accretion onto a class of BHs and found that CCH
is valid for phantom accretion onto a stringy charged BH.

Motivated by the recent development in Horava-Lifshitz gravity, we
investigate the phantom accretion onto a static Kehagias-Sfetsos
(KS) BH in Ho$\check{r}$ava Lifshitz (HL) gravity, by using
Babichev-Dokuchaev-Eroshenko method [17] and discuss the locations
of the critical points of accretion. Further, the relations between
critical points and horizons have been found. The gravitational
units are used. All the Latin and Greek indices vary from 0 to 3,
otherwise it will be stated.

\section{Phantom Energy Accretion Onto a BH in Horava-Lifshitz Gravity }

The KS BH solution [17] is given by
\begin{equation}\label{1}
ds^2=A(r)dt^2-\frac{1}{A(r)}dr^2-r^2(d\theta^2+\sin^2\theta d\phi^2)
\end{equation}
where $A(r)=1+wr^2-wr^2\sqrt{{1+\frac{4m}{wr^3}}}$ and
$w=\frac{16\mu^2}{{\kappa}^2}$ ($\mu$ and $\kappa$ are constants),
for $w\rightarrow\infty$, KS BH $\rightarrow$ Schwarzschild BH. The
horizons of KS BH can be obtained by solving $A(r)=0$, for $r$.
Hence, in this case horizons are given by
\begin{equation}\label{2}
r_{\pm}=m\left(1\pm \sqrt{{1-\frac{1}{2wm^2}}}\right).
\end{equation}

The energy momentum tensor representing phantom energy in perfect
fluid given by
\begin{equation}\label{7}
{T_{{\mu}{\nu}}={({\rho}+p)}u_{\mu}u_{\nu}-pg_{\mu\nu}},
\end{equation}
where $\rho$ and $p$ are energy density and pressure of phantom
energy and $u^\mu=(u^t,u^r,0,0)$ is the velocity four-vector of
fluid flow. We note that $u^\mu$ satisfies the normalization
condition, that is $u^\mu u_\mu =1$.

For the phantom energy accretion onto a KS BH, we shall drive two
equations of motions, one by the conservation of energy-momentum
tensor and other by projecting the energy-momentum conservation law
on the four-velocity. The energy conservation equation
$T^{0\mu}_{;\mu}=0$ is given by
\begin{equation}
\label{8}
r^2u(\rho+p)\left(1+wr^2-wr^2\sqrt{{1+\frac{4m}{wr^3}}}+u^2\right)^{\frac{1}{2}}=B_0,
\end{equation}
where $B_0$ is an integration constant and $u^{r}=u<0$ for the
inward phantom flow. Further, the energy flux equation can be
derived by projecting the energy-momentum conservation law on the
four-velocity, that is, ${u_\mu T^{\mu\nu}}_{;\nu}$=0 for which
Eq.(\ref{7}) leads to
\begin{equation}\label{9}
r^2u\exp
\left[\int^\rho_{\rho_\infty}\frac{d\rho'}{\rho'+p(\rho')}\right]=-B_1,
\end{equation}
where $B_1>0$ is another integration constant which is related to
the energy flux. Also, ${\rho}$ and ${\rho_\infty}$ are densities of
the phantom energy at finite and infinite $r$. From Eqs.(\ref{8})
and (\ref{9}), we can obtain
\begin{equation}\label{10}
(\rho+p)\left(1+wr^2-wr^2\sqrt{{1+\frac{4m}{wr^3}}}+u^2\right)^{\frac{1}{2}}
\exp\left[-\int^\rho_{\rho_\infty}\frac{d\rho'}{\rho'+p(\rho')}\right]=B_2,
\end{equation}
where $B_2=-\frac{B_0}{B_1}=\rho_\infty +p(\rho_\infty)$.

The rate of change of BH mass due to phantom accretion is given by
[12]
\begin{equation}\label{11}
\dot{m}=4 \pi r^2 {T^r}_0.
\end{equation}
Using Eqs.(\ref{9}) and (\ref{10}) in the above equation yields
\begin{equation}\label{12}
\dot{m}=4 \pi B_1 ({\rho}_\infty +{p}_\infty).
\end{equation}
We note that the mass of BH decreases if $({\rho}_\infty
+{p}_\infty)<0$. Thus the accretion of phantom energy onto a BH
decreases the mass of BH. It can be noted here that one can solve
Eq.(\ref{12}) for $m$ by using EoS $p=\omega\rho$. Since all $p$ and
$\rho$, violating dominant energy condition, must satisfy this
equation, hence it holds in general, that is
\begin{equation}\label{12a}
\dot{m}=4 \pi B_1 ({\rho} +{p}).
\end{equation}.

Now, we analyze the critical points (such points at which flow speed
is equal to the speed of sound) during the accretion of phantom
energy. The phantom energy falls onto BH with increasing velocity
along the particle trajectories. The conservation of mass flux is
\begin{equation}\label{13}
\rho u r^2=D_0,
\end{equation}
where $D_0$ is the constant of integration. Dividing and squaring
Eqs.(\ref{8}) and (\ref{13}), we obtain
\begin{equation}\label{14}
\left(\frac{\rho +p}{\rho}\right)^2 \left(
1+wr^2-wr^2\sqrt{{1+\frac{4m}{wr^3}}}+u^2\right)=D_1,
\end{equation}
where $D_1=(\frac{B_0}{D_0})^2$ is a positive constant.
Differentiating Eqs.(\ref{13}) and (\ref{14}) and eliminating
$d\rho$, we derive
\begin{equation}\label{15}
\frac{dr}{r}\left[2V^2-\frac{wr^2-wr^2\sqrt{{1+\frac{4m}{wr^3}}}+\frac{6m}{r}(1+\frac{4m}{wr^3})^{\frac{-1}{2}}}
{A(r)+u^2}\right]+\frac{du}{u}
\left[V^2-\frac{u^2}{A(r)+u^2}\right]=0.
\end{equation}
where $V^2=\frac{d\ln(\rho+p)}{d\ln\rho}-1$ and $A(r)$ is the lapse
function as defined after Eq.(\ref{1}).

This equation shows that turn-around points (critical points) are
located where both the square brackets vanish. Thus
\begin{eqnarray}\label{16}
{V_c}^2&=&\frac{wr^2-wr^2\sqrt{{1+\frac{4m}{wr^3}}}+\frac{3m}{r}(1+\frac{4m}{wr^3})^{\frac{-1}{2}}}
{2(A(r)+{u_c}^2)} ,\\{V_c}^2&=&\frac{{u_c}^2}{A(r)+{u_c}^2}.
\end{eqnarray}
Solving above equations for ${u_c}^2$ and ${V_c}^2$, we can obtain
\begin{eqnarray}\label{16a}
{u_c}^2&=&\frac{1}{2}\left(wr^2-wr^2\sqrt{{1+\frac{4m}{wr^3}}}+\frac{3m}{r}(1+\frac{4m}{wr^3})^{\frac{-1}{2}}\right),\\\label{16b}
{V_c}^2&=&\frac{\left(wr^2-wr^2\sqrt{{1+\frac{4m}{wr^3}}}+\frac{3m}{r}(1+\frac{4m}{wr^3})^{\frac{-1}{2}}\right)}{2+3
\left(wr^2-wr^2\sqrt{{1+\frac{4m}{wr^3}}}+\frac{m}{r}(1+\frac{4m}{wr^3})^{\frac{-1}{2}}\right)}.
\end{eqnarray}
We can note that in the limit $w\rightarrow\infty$, the above
equations reduce to ${u_c}^2=\frac{m}{2r_c}$ and
${V_c}^2=\frac{m}{2r_c-3m}$, that is the Schwarzschild HB case [9].
The physically acceptable solutions of Eqs.(\ref{16a}) and
(\ref{16b}) are obtained if ${u_c}^2>0$ and ${V_c}^2>0$, implying
that
\begin{figure}
\epsfig{file=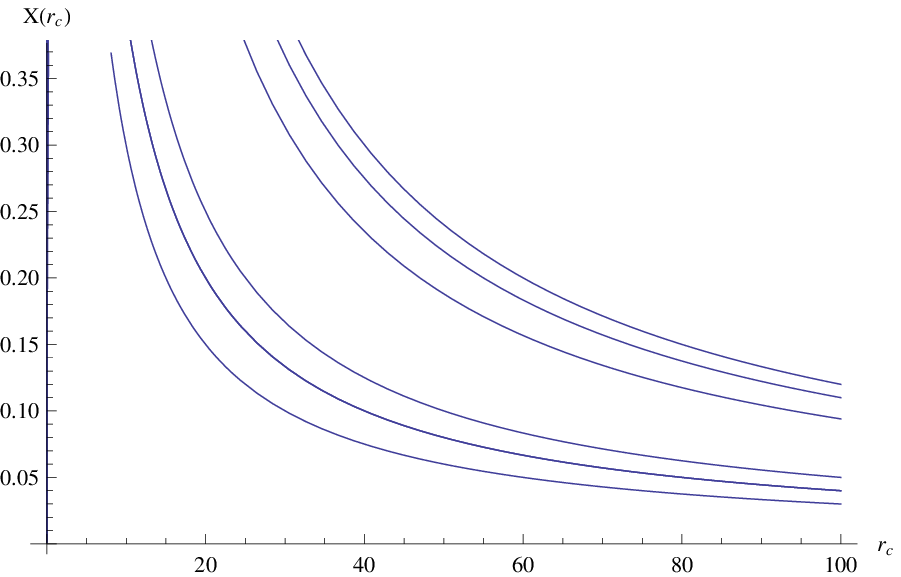, width=0.5\linewidth}\epsfig{file=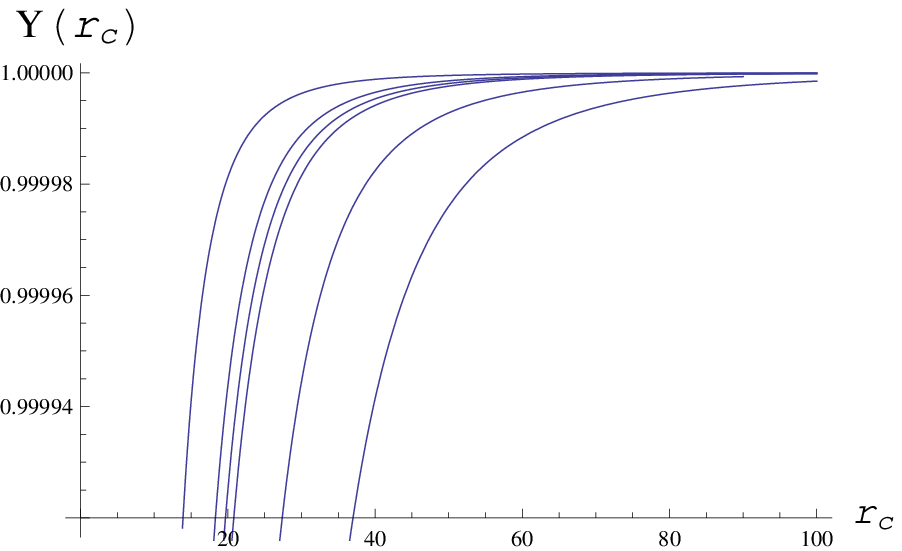,
width=0.5\linewidth} \caption{The left right graphs represent the
behavior of $X(r_c)$ and $Y(r_c)$, respectively. In both graphs the
curves of $X(r_c)$ and $Y(r_c)$ have been plotted for
$w=m=1,2,3,4,5,50$ corresponding to up to down order of curves}
\end{figure}
\begin{eqnarray}\label{18}
X(r)=\frac{1}{2}\left(wr^2-wr^2\sqrt{{1+\frac{4m}{wr^3}}}+\frac{3m}{r}(1+\frac{4m}{wr^3})^{\frac{-1}{2}}\right)>0,\\
\label{19}
Y(r)=2+{3}\left(wr^2-wr^2\sqrt{{1+\frac{4m}{wr^3}}}+\frac{m}{r}(1+\frac{4m}{wr^3})^{\frac{-1}{2}}\right)>0.
\end{eqnarray}
It is not possible to determine analytically the values of $w, m$
and $r_c$, for which $X(r_c)>0$ and $Y(r_c)>0$, however we can
obtain the positivity of these quantities graphically as shown in
Fig. \textbf{{1}}.

\section{Summary}

It is known that a BH surrounding the matter is expected to capture
particles of a matter that passes near BH. In this paper, we have
explored the phantom accretion onto a BH in HL gravity. By assuming
that infalling fluid does not alter the generic properties of the
BH, the equations of motion for steady state spherically symmetric
phantom flow near BH have been derived. We herein discuss the
accretion and critical accretion onto a BH by using
Babichev-Dokuchaev-Eroshenko method [10]. It has been found that
like the cases of the Schwarzschild and RN BHs, phantom accretion
decreases the mass of BH in HL gravity.
\begin{table}[ht]
\caption{Location of horizons for the values of $m$ and $w$}
\begin{center}
\begin{tabular}{|c|c|c|c|c|}
\hline \textbf{$m=w$}&  \textbf{ $r_-$} & \textbf{$r_+$}
\\\hline  1& 0.2928   &
1.7072
\\\hline 2& 0.1291&
$3.8708$
\\\hline 3&0.0845&
$5.9145$
\\\hline 4&0.0629&
$7.9370$
\\\hline 5&0.0146&
$9.9853$
\\\hline 6&0.0839&
$11.9581$
\\\hline 7&0.0358&
$13.9642$
\\\hline 8&0.0313&
$15.9686$
\\\hline 50&0.0018&
$99.9949$
\\\hline
\end{tabular}
\end{center}
\end{table}
There exist two horizons for the HL BH, which depends on $m$ and
$w$. The critical accretion analysis implies that accreting fluid
would attain speed of sound if $1\leq m=w\leq 50$. The quantities
$X(r)$ and $Y(r)$ plotted in Fig. \textbf{1}, imply that for $1\leq
m=w\leq 50$, the critical accretion is possible as
$V^2_c\geqslant0$. The location of critical point in case is
$15\leqslant r_c\leqslant 100$. Thus the location of inner and outer
horizons (Eq.(\ref{2})) for $1\leq m=w\leq 50$ is given by Table
\textbf{1}.

From the Fig. \textbf{1} and Table \textbf{1} , we observe that for
$1\leq m=w\leqslant7$, the critical accretion points lie outside the
inner and outer horizons. For $7<m=w<50$, the critical accretion
pints lies inside the outer horizons but outside inner horizons.
Thus, we conclude that for each value of $1\leq m=w\leqslant7$,
there exist two circular horizons (inner and outer) that are bounded
by a circle of larger radius, representing the critical accretion
region. As $m$ becomes large $\leqslant50$ the outer horizon lies
into the critical accretion region.

\vspace{1.0cm}

1~~~~Perlmutter S, et al. Measurements of cosmological parameters $\Omega$ and

~~~~~ $\Lambda$ from the first seven Supernovae at $z\geq 0.35$.Astrophys J,

~~~~ 9997, 483: 565-581\\

2~~~~Perlmutter S, et al. Discovery of supernova explosion at half the

~~~~age of Universe. Nature, 1998, 391: 51-54\\

3~~~~Perlmutter S, et al.  Measurements of Omega and Lambda from 42

~~~~~high-redshift Supernovae. Astrophys J, 1999, 517: 565-586\\

4~~~~Riess A G,  et al. Observational evidence from Supernovae for

~~~~accelerating Universe and cosmological constant. Astron J, 1998,

~~~~116: 1009-10038\\

5~~~~Bennett C L, et al. First-Year Wilkinson Microwave Anisotropy

~~~~Probe (WMAP) observations: Preliminary and basic results.

~~~~Astrophys J Suppl, 2003, 148: 1-27\\

6~~~~Spergel D N, et al. First-Year Wilkinson Microwave Anisotropy

~~~~Probe (WMAP) observations: Determination of cosmological

~~~~parameters. Astrophys J Suppl, 2003, 148: 175-194\\

7~~~~Verde L, et al. The 2dF galaxy redshift survey: The bias of the

~~~~galaxies and the density of the universe. Mon Not R Astron Soc,

~~~~2002, 335: 432-440\\

8~~~~Bondi H. On Spherically symmetric accretion. Mon Not Roy Astron

~~~~Soc, 1952, 112: 195-204\\

9~~~~Michel F C. Accretion of matter by condensed objects. Astrophys

~~~~Space Sci, 1972, 15: 153-160\\

10~~~~Babichev E, Dokuchaev V, Eroshenko Y. Black hole mass

~~~~decreasing due to phantom energy accretion. Phys Rev Lett, 2004,

~~~~~~~ 93: 021102-021105\\

11~~~~ Jamil M, Rashid M, Qadir A. Charged black hole in phantom

~~~~cosmology. Eur Phys J C, 2008, 58: 325-329\\

12~~~~ Babichev E, Dokuchaev V, Eroshenko Y. Perfect fluid and

~~~~ Scalar field in Reissner-Nordstrom metric. J Exp Theor Phys,

~~~~2011, 112: 784-793\\

13~~~~ Jimenez Madrid J A, Gonzalez-Dias P F. Evaluation of

 ~~~~ Kerr-Newman black hole in a dark energy universe. Gravit Cosmol,

~~~~~~2008,14: 213-225\\

14~~~~Sharif M, Abbas G. Phantom accretion onto the Schwarzschild

~~~~black hole. Chin Phys Lett, 2011, 28: 090402-4\\

15~~~~Sharif M, Abbas G. Phantom energy accretion by a string

~~~~ charged black hole. Chin Phys Lett, 2012, 29: 010401-3\\

16~~~~Sharif M, Abbas G. Phantom energy accretion by a class of

~~~~ black holes. J Phys Conf Ser, 2012, 354: 012019-9\\

17~~~~Liu M, Lu J.  Logarithmic entropy of Kehagias-Sfetsos black

~~~~ hole with self gravitation in asymptotically flat FRW IR

~~~~ modified Horava gravity. Phys Lett B, 2011, 699: 296-300
\end{document}